\documentclass[preprint,prd,
superscriptaddress,nofootinbib,showpacs,tightenlines]{revtex4-1}
\usepackage{amssymb,latexsym}
\usepackage{amsmath,amsbsy}
\usepackage{epsfig,bm}
\usepackage{graphicx}
\usepackage{comment}
\DeclareMathOperator{\tr}{tr}

\DeclareFontFamily{OT1}{pzc}{}
\DeclareFontShape{OT1}{pzc}{m}{it}%
             {<-> s * [1.40] pzcmi7t}{}
\DeclareMathAlphabet{\mathscr}{OT1}{pzc}%
                                 {m}{it}

\begin{document}
\def\a{{\alpha}}
\def\b{{\beta}}
\def\d{{\delta}}
\def\D{{\Delta}}
\def\e{{\varepsilon}}
\def\g{{\gamma}}
\def\G{{\Gamma}}
\def\k{{\kappa}}
\def\l{{\lambda}}
\def\L{{\Lambda}}
\def\m{{\mu}}
\def\n{{\nu}}
\def\o{{\omega}}
\def\O{{\Omega}}
\def\S{{\Sigma}}
\def\s{{\sigma}}
\def\th{{\theta}}

\def\ol#1{{\overline{#1}}}

\def\Dslash{D\hskip-0.65em /}
\def\Dtslash{\tilde{D} \hskip-0.65em /}

\def\CPT{{$\chi$PT}}
\def\QCPT{{Q$\chi$PT}}
\def\PQCPT{{PQ$\chi$PT}}
\def\tr{\text{tr}}
\def\str{\text{str}}
\def\diag{\text{diag}}
\def\order{{\mathcal O}}

\def\meff{{m^2_{\text{eff}}}}

\def\Meff{{M_{\text{eff}}}}
\def\cF{{\mathcal F}}
\def\cS{{\mathcal S}}
\def\cC{{\mathcal C}}
\def\cE{{\mathcal E}}
\def\cB{{\mathcal B}}
\def\cT{{\mathcal T}}
\def\cQ{{\mathcal Q}}
\def\cL{{\mathcal L}}
\def\cO{{\mathcal O}}
\def\cA{{\mathcal A}}
\def\cV{{\mathcal V}}
\def\cR{{\mathcal R}}
\def\cH{{\mathcal H}}
\def\cW{{\mathcal W}}
\def\cM{{\mathcal M}}
\def\cD{{\mathcal D}}
\def\cN{{\mathcal N}}
\def\cP{{\mathcal P}}
\def\cK{{\mathcal K}}
\def\Qt{{\tilde{Q}}}
\def\Dt{{\tilde{D}}}
\def\psit{{\tilde{\psi}}}
\def\St{{\tilde{\Sigma}}}
\def\cBt{{\tilde{\mathcal{B}}}}
\def\cDt{{\tilde{\mathcal{D}}}}
\def\cTt{{\tilde{\mathcal{T}}}}
\def\cMt{{\tilde{\mathcal{M}}}}
\def\At{{\tilde{A}}}
\def\Qt{{\tilde{Q}}}
\def\cNt{{\tilde{\mathcal{N}}}}
\def\cOt{{\tilde{\mathcal{O}}}}
\def\cPt{{\tilde{\mathcal{P}}}}
\def\cI{{\mathcal{I}}}
\def\cJ{{\mathcal{J}}}

\def\eqref#1{{(\ref{#1})}}

\preprint{MIT-CTP-4090}
\preprint{UMD-40762-473}

\title{Kaon Thresholds and Two-Flavor Chiral Expansions for Hyperons}

\author{F.-J.~Jiang}
\email[]{fjjiang@ntnu.edu.tw}
\affiliation{%
Center for Theoretical Physics,
Department of Physics, 
Massachusetts Institute of Technology, 
Cambridge, MA 02139,
USA
}
\affiliation{%
Department of Physics,
National Taiwan Normal University, 
88, Sec.~4, Ting-Chou Rd., 
Taipei 116,
Taiwan
}

\author{B.~C.~Tiburzi}
\email[]{bctiburz@umd.edu}
\affiliation{%
Maryland Center for Fundamental Physics, 
Department of Physics, 
University of Maryland, 
College Park,  
MD 20742-4111, 
USA
}

\author{A.~Walker-Loud}
\email[]{walkloud@wm.edu}
\affiliation{%
Department of Physics, 
College of William and Mary, 
Williamsburg, VA 23187-8795,
USA
}

\date{\today}


\begin{abstract}
Two-flavor chiral expansions provide a useful perturbative framework to study hadron properties. 
Such expansions should exhibit marked improvement over the conventional three-flavor chiral expansion. 
Although one can theoretically formulate two-flavor theories for the various hyperon multiplets, 
the nearness of kaon thresholds can seriously undermine the effectiveness of the perturbative expansion in practice. 
We investigate the importance of virtual kaon thresholds on hyperon properties, 
specifically their masses and isovector axial charges. 
Using a three-flavor expansion that includes 
$SU(3)$ 
breaking effects,  
we uncover the underlying expansion parameter governing the description of virtual kaon thresholds. 
For spin-half hyperons, 
this expansion parameter is quite small. 
Consequently
virtual kaon contributions are well described in the two-flavor theory by terms analytic in the pion mass-squared. 
For spin three-half hyperons,
however, 
one is closer to the kaon production threshold, 
and the expansion parameter is not as small. 
Breakdown of 
$SU(2)$
chiral perturbation theory is shown to arise from a pole in the expansion parameter associated with the kaon threshold.
Estimating higher-order corrections to the expansion parameter 
is necessary to ascertain whether the two-flavor theory of spin three-half hyperons
remains perturbative. 
We find that, 
despite higher-order corrections, 
there is a useful perturbative expansion for the masses and isovector axial charges of both spin-half and spin three-half hyperons. 
\end{abstract}
\maketitle

\section{Introduction}
\label{what?}

The low-energy regime of QCD can be described by an effective field theory. 
This theory, 
chiral perturbation theory, 
encodes the pattern of spontaneous and explicit chiral symmetry breaking present in QCD. 
Using chiral perturbation theory, 
hadron properties can be determined in terms of universal low-energy parameters in 
an expansion about vanishing light quark masses. 
The three-flavor chiral expansion relies upon treating the up, 
down, 
and strange quark masses as small compared to the QCD scale. 
In the baryon sector, 
this expansion has well-known convergence issues, 
and attempts have been made to improve the expansion, 
see, e.g.,~%
\cite{Donoghue:1998bs}.

The physical mass of the strange quark is potentially too large to be considered a small perturbation about the 
$SU(3)$ 
chiral limit. 
Phenomenological and lattice QCD calculations have determined the ratio of light quark masses to be
$m_s / \hat{m} \sim 25$, 
where 
$m_s$ 
is the strange quark mass, 
and 
$\hat{m}$ 
is the average of up and down quark masses. 
The size of this ratio suggests an alternate expansion: 
treat only the lighter up and down quark masses as small,
and expand about the 
$SU(2)$ 
chiral limit. 
This approach has long been advocated for pions~%
\cite{Gasser:1983yg}, 
nucleons~%
\cite{Bernard:1995dp}, 
and deltas~%
\cite{Hemmert:1997ye}, 
however, 
only recently have strange hadrons been treated in the two-flavor chiral expansion. 
Following earlier work of~%
\cite{Roessl:1999iu},
kaon properties have been investigated in 
$SU(2)$ 
chiral perturbation theory~%
\cite{Allton:2008pn,Flynn:2008tg,Bijnens:2009yr}.
This renewed interest stems from lattice QCD applications. 
Current simulations no longer require extrapolation in the strange quark mass, 
rather an interpolation. 
Consequently formulae parametrizing only the pion mass dependence of observables are required, 
for which 
$SU(2)$
is an ideal framework,
independent of the potential convergence issues of 
$SU(3)$. 
Baryons with non-vanishing strangeness have been treated using 
$SU(2)$
chiral perturbation theory~%
\cite{Frink:2002ht,Beane:2003yx,Tiburzi:2008bk,Jiang:2009sf,Mai:2009ce,Jiang:2009jn}.
Much of this work, too, has been motivated by progress in lattice QCD computations.

As with any effective theory, 
it is not \emph{a priori} obvious that an 
$SU(2)$ 
treatment of hyperons is possible.
Consider the 
$\S$ 
baryons. 
In order to describe the 
$\S$ 
multiplet in 
$SU(2)$
chiral perturbation theory, 
the 
$N$ 
and 
$\Xi$ 
multiplets must be energetically well separated.
This separation occurs in nature due to the size of the strange quark mass. 
If the separation were much larger, 
however, 
the 
$\S$ 
would decay strongly to 
$K N$. 
Thus the baryon mass splittings must be large compared to the pion mass;
but na\"ively small compared to the kaon mass. 
While all hyperons are stable with respect to strong strangeness-changing decays, 
not all baryon mass splittings are small compared to the kaon mass. 
A natural question emerges:
without explicit kaons, 
can the 
$SU(2)$
expansion reproduce the non-analyticities required sub-threshold?
We observe that it appears to be possible; 
however, 
we cannot answer the question
for all low-energy hyperon observables.
This observation was alluded to in~%
\cite{Tiburzi:2009ab}, 
and our goal is to concretely solidify the argument. 
As we consider only the specific examples of hyperon masses and isovector axial charges, 
further work is needed to clarify when an 
$SU(2)$
treatment is warranted by nature. 
The efficacy is quite likely observable dependent.

We employ the following organization for our presentation. 
First in Sec.~\ref{twoflavor}, 
we motivate the two-flavor chiral expansion through the investigation of hyperon masses. 
We begin by considering the
$SU(3)$ 
symmetric case, 
and then proceed to include 
$SU(3)$
breaking,
which is necessary to account for virtual kaon production thresholds. 
We deduce the expansion parameter that controls the effects
of kaon thresholds in the two-flavor theory. 
We then investigate how well the kaon threshold is reproduced in 
$SU(2)$ 
chiral perturbation theory in Sec.~\ref{s:effect}.
Specifically 
we focus on the kaon-baryon sunset diagrams that contribute to hyperon masses, 
and isovector axial charges. 
A brief summary,
Sec.~\ref{summy}, 
concludes this work. 
Finally, 
the Appendix is devoted to estimating higher-order corrections to expansion parameters.

\section{Two-Flavor Chiral Expansion for Hyperons}
\label{twoflavor}                                                                                             %

In order to investigate the effect of virtual kaon thresholds on hyperon properties, 
we begin with 
$SU(3)$
heavy baryon chiral perturbation theory~\cite{Jenkins:1990jv,Jenkins:1991es}. 
Dissecting an explicit kaon loop contribution, 
we uncover the parameters governing the convergence of the two-flavor expansion. 
This requires 
$SU(3)$
breaking corrections.

\subsection{Schematic Example}
\label{ss:se}

Kaon and eta loops typically yield large numerical contributions to baryon observables in 
$SU(3)$
chiral perturbation theory. 
For this reason, 
it is efficacious to have an 
$SU(2)$
expansion of baryon properties; 
so that, with only explicit pions,
the convergence properties of the theory are markedly improved. 
In~\cite{Tiburzi:2008bk}, 
the 
$SU(2)$ 
theory of hyperons was written down by appealing to symmetries
that emerge when the quark masses are treated in the hierarchy
\begin{equation} \label{eq:h2}
\hat{m} \ll m_s \sim \L_{QCD}
.\end{equation}
As a consequence, 
the efficacy of this theory is determined by the size of the quark mass ratio, 
$\varepsilon_{SU(2)} \equiv \hat{m} / m_s \sim 1 / 25$. 
This estimate is na\"ive, 
however, 
because it cannot account for non-perturbative contributions. 
A way to infer the underlying expansion parameters is to match 
$SU(3)$
onto  
$SU(2)$.

As the fate of the 
$SU(2)$ 
expansion for hyperons is largely determined by the nearness of kaon thresholds, 
we focus on the kaon mass, 
$m_K$, 
and the $m_K$-dependence of kaon loop contributions.
Using the hierarchy in Eq.~\eqref{eq:h2},
we can expand the kaon mass about 
$\hat{m} = 0$,
namely%
~\cite{Roessl:1999iu}
\begin{equation} \label{eq:kaon}
m_K^2 
= 
[m_K^{SU(2)}]^2 +
\frac{1}{2} C \,
m_\pi^2 
+ 
\mathcal{O} ( \varepsilon^2_{SU(2)} )
,\end{equation}
where 
$m_K^{SU(2)}$
is the kaon mass in the 
$SU(2)$ 
chiral limit,
and 
$C$ 
is a low-energy constant of 
$SU(2)$
chiral perturbation theory. 
This low-energy constant depends on 
$m_K^{SU(2)}$
in the form
$C = C \left( [m_K^{SU(2)}]^2 / \L^2_\chi \right)$, 
where
$\L_\chi \approx 1.1 \, \texttt{GeV}$ 
is the chiral symmetry breaking scale.
To estimate
$m_K^{SU(2)}$,
we appeal to 
$SU(3)$ 
chiral perturbation theory. 
At leading order, 
the Gell-Mann--Oakes--Renner (GMOR) relation implies the value
$C(0) = 1$. 
Using the neutral pion mass and the average mass-squared of the kaons, 
we find 
\begin{equation} \label{eq:kaon2}
m_K^{SU(2)} = 0.486(5) \, \texttt{GeV}
,\end{equation}
where the uncertainty has been estimated from the analytic term of
$\mathcal{O}(x)$
in the expansion of 
$C(x)$ 
about 
$x =0$,
assuming the value
$C'(0) = 2$. 
The 
$\mathcal{O}(x)$ 
corrections including logarithms are known from 
$SU(3)$ 
chiral perturbation theory, 
but depend on low-energy constants that are not precisely determined from phenomenology. 
In the Appendix, 
we use phenomenological and lattice QCD inputs to estimate    
$m_K^{SU(2)}$
at next-to-leading order in 
$SU(3)$,
and find that all estimates lie within the error-bar quoted in Eq.~\eqref{eq:kaon2}.

Now we turn to kaon loop contributions to hyperon observables.
As a schematic example, 
we consider the mass of the 
$\S$
baryon. 
In 
$SU(3)$
chiral perturbation theory, 
the leading kaon loop contribution enters at third order,
$\mathcal{O} (m_K^3)$.
Writing only this contribution, 
we have
\begin{equation} \label{eq:sigma}
\D M_\S = a_K \, m_K^3
,\end{equation}
where 
$a_K$
depends on the low-energy constants of 
$SU(3)$. 
Inserting the 
$SU(2)$ 
expansion of the kaon mass from Eq.~\eqref{eq:kaon} into the loop contribution in Eq.~\eqref{eq:sigma}, 
we find
\begin{equation} \label{eq:sigma2}
\D M_\S 
= 
a_K 
[m_K^{SU(2)} ]^3
+ 
\frac{3}{4} a_K \, C \, m_K^{SU(2)}  \,\, m_\pi^2
+ 
\mathcal{O} ( \varepsilon_{SU(2)}^2 )
.\end{equation}
Above, 
the first term is a contribution to the 
$\S$ 
mass in the  
$SU(2)$
chiral limit, 
$M_\S^{SU(2)}$,
while the second term 
is a contribution to the 
$\pi$-$\S$ 
sigma term of 
$SU(2)$, 
which has a form
$\propto \sigma_\S \, m_\pi^2$. 
In carrying out the 
$SU(2)$ 
expansion,
non-analytic kaon mass dependence is
traded for a tower of terms analytic in the pion mass squared. 
The only non-analytic pion mass dependence arises from pion loops. 
The convergence of the $\S$ mass in 
$SU(2)$ is governed by: 
the chiral expansion, 
$m_\pi^2 / \Lambda_\chi^2$;
and the heavy $\S$ expansion,
$m_\pi / M_\S^{SU(2)}$. 
This reorganization is possible due to the small parameter
$\varepsilon_{SU(2)}$
that underlies the 
$SU(2)$
expansion of kaon contributions%
\footnote{
The small parameter
$\frac{1}{2} \varepsilon_{SU(2)}$ 
underlies the expansion of eta loop contributions.}.

We have presented a schematic argument to motivate the 
$SU(2)$ 
expansion of the 
$\S$ 
mass. 
This argument generalizes to other hyperons and to other observables; 
however, 
we have ignored baryon mass splittings in loop contributions. 
These require a more careful treatment, 
to which we now turn.

\subsection{Kaon Production Thresholds}
\label{s:KPT}

An expansion of hyperon observables in powers of 
$\varepsilon_{SU(2)}$  
is very well behaved. 
There are additional expansion parameters, 
however, 
that underly the 
$SU(2)$ 
theory of hyperons. 
These additional parameters are related to kaon production thresholds.
Clearly for the two-flavor theory to be effective, 
kaon production thresholds cannot be reached. 
When this condition is met, 
the kaons and eta need not appear explicitly in the effective theory, 
and their virtual loop contributions can be reordered.
Such an 
$SU(2)$ 
formulation can describe the virtual strangeness changing transitions
provided one is suitably far from thresholds. 
We make this criterion quantitative by considering the effect of kaon production thresholds
in the matching of
$SU(3)$ 
onto 
$SU(2)$.

Let us return to 
$SU(3)$
chiral perturbation theory. 
Loop diagrams in which the baryon strangeness changes typically have non-negligible mass splittings 
between the external and intermediate-state baryons. 
This is primarily due to the strange quark mass, 
and it is efficacious for the two-flavor expansion to consider these baryon mass splittings in the 
$SU(2)$ 
chiral limit. 
For example, 
a generic $B' \to K B$ process
is a $\D S = -1$ strangeness changing baryon transition, 
and is characterized by the mass splitting
$\delta_{BB'}$, 
given by
\begin{equation}
\delta_{BB'} = M_{B'}^{SU(2)} - M_{B}^{SU(2)}
.\end{equation}
When the physical mass splitting exceeds the kaon mass, 
the decay is kinematically allowed, 
otherwise the process 
$B' \to K B$ is virtual.

To estimate the 
$SU(2)$ 
chiral limit splittings, 
we use the experimental values for the isospin averaged baryon masses. 
This is a leading-order estimate, 
and higher-order corrections are considered in the Appendix. 
Values for the positive
$\D S = - 1$
splittings between baryons are: 
$\delta_{N \Sigma^*} = 0.45 \, \texttt{GeV}$, 
$\delta_{\Lambda \Xi^*} = 0.42 \, \texttt{GeV}$, 
$\delta_{\Xi \Omega} = 0.36 \, \texttt{GeV}$, 
$\delta_{\Sigma \Xi^*} = 0.34 \, \texttt{GeV}$, 
$\delta_{N \Sigma} = 0.25 \, \texttt{GeV}$, 
$\delta_{\Lambda \Xi} = 0.20 \, \texttt{GeV}$, 
$\delta_{N \Lambda} = 0.18 \, \texttt{GeV}$, 
$\delta_{\Delta \Sigma^*} = 0.15 \, \texttt{GeV}$, 
$\delta_{\Sigma^* \Xi^*} = 0.15 \, \texttt{GeV}$, 
$\delta_{\Xi^* \Omega} = 0.14 \, \texttt{GeV}$, 
and
$\delta_{\Sigma \Xi} = 0.12 \, \texttt{GeV}$,
while there are a few positive 
$\D S = 1$
splittings as well: 
$\delta_{\Lambda \D} = 0.12 \, \texttt{GeV}$, 
$\delta_{\Xi \Sigma^*} = 0.07 \, \texttt{GeV}$, 
and
$\delta_{\Sigma \D} = 0.04 \, \texttt{GeV}$. 
The latter describe processes of the generic form 
$B' \to \ol K B$. 
While all 
$|\Delta S| = 1$
splittings are below threshold, 
$\delta_{BB'} < m_K$, 
with 
$m_K = 0.50 \, \texttt{GeV}$,
certain spin-3/2 to spin-1/2 transitions are not considerably far from threshold. 
At first glance, 
it appears that the 
$SU(2)$
theory will poorly describe the non-analyticities associated with such inelastic thresholds. 
This impression is based on the value of 
$ \delta_{BB'} / m_K \sim 1$; 
which, 
however, 
is not the appropriate expansion parameter for 
$SU(2)$
chiral perturbation theory.

To deduce the expansion parameter relevant for an 
$SU(2)$ 
description of hyperons, 
we focus on a schematic example, 
and include the splitting, 
$\delta_{BB'}$. 
The introduction of this scale into loop integrals produces a more complicated non-analytic function involving both 
$m_K$ 
and 
$\d_{BB'}$. 
For diagrams of the sunset type, 
a logarithm is generically produced of the form
\begin{equation}
\cL (m_K^2, - \d_{BB'} )
= 
\log 
\left( 
\frac{- \d_{BB'} - \sqrt{\d_{BB'}^2 - m_K^2 + i \epsilon}}{ - \d_{BB'} + \sqrt{\d_{BB'}^2 - m_K^2 + i \epsilon}}
\right)
,\end{equation}
which contains the non-analyticities associated with kaon production. 
We stress that this example is schematic. 
Explicit functions describing loop contributions to masses and axial charges will be considered in detail below. 
Just above threshold, 
$\delta_{BB'}  \gtrsim m_K$, 
the logarithm behaves as
\begin{equation}
\mathcal{L}(m_K^2, - \delta_{BB'})  
\longrightarrow 
- 2 \pi i  + \ldots
.\end{equation}
The imaginary part of the logarithm is associated with the width for the 
real decay process 
$B' \to K \, B$.
In this regime, 
an 
$SU(2)$
description fails as
explicit kaon degrees of freedom are required.

For the mass splittings listed above, 
however,
our concern is with the region below threshold. 
When 
$\delta_{BB'}  \lesssim m_K$, 
the 
$SU(2)$ 
treatment must also fail, 
and we address whether the physical splittings actually put us in this regime. 
Applying the perturbative expansion about the 
$SU(2)$ 
chiral limit for the generic logarithm, 
we make the following observation:
terms in the logarithm that are expanded can be written as functions of the form 
\begin{equation} \label{eq:expand}
f \Big( m_K^2 - \delta_{BB'}^2 \Big)
=
f \Big( [m_K^{SU(2)}]^2 - \delta_{BB'}^2 \Big)
+
\left( m_K^2 - [m_K^{SU(2)}]^2 \right)
f' \Big( [m_K^{SU(2)}]^2  - \delta_{BB'}^2 \Big)
+
\ldots \,
.\end{equation}
Thus for the subthreshold case, 
the expansion parameter, 
$\varepsilon_{BB'} $,
is of the form
\begin{equation} \label{eq:discovery}
\varepsilon_{BB'} 
= 
\frac{\frac{1}{2} C \, m_\pi^2}{ [m_K^{SU(2)}]^2 -  \delta_{BB'}^2}
,\end{equation}
having dropped terms of 
$\mathcal{O} ( \varepsilon_{SU(2)} )$. 
When the baryon mass splitting is negligible compared to the chiral limit kaon mass,
$\delta_{BB'} \ll m_K^{SU(2)}$, 
we arrive at 
$\varepsilon_{BB'} = \varepsilon_{SU(2)}$
by utilizing the GMOR relation to set 
$C =1$. 
This reduces to the case considered above in 
Sec.~\ref{ss:se}, 
where we neglected the baryon mass splittings. 
On the other hand, 
in the limit 
$\delta_{BB'} \to m_K^{SU(2)}$,
the expansion parameter becomes arbitrarily large. 
This is the signal of the breakdown of 
$SU(2)$
chiral perturbation theory.
With 
$\varepsilon_{BB'} \sim 1$, 
non-analyticities associated with the virtual kaon threshold 
cannot be described by a perturbative expansion in the pion mass-squared.

For the strangeness transitions listed above, 
we can diagnose the convergence properties of the
$SU(2)$
expansion by estimating the size of the expansion parameters governing
the description of kaon thresholds. 
We use the leading-order values for the masses along with 
$C = 1$;
higher-order corrections are discussed in the Appendix.  
For the 
$\D S = - 1$ 
virtual transitions, 
we have:
$\varepsilon_{N \Sigma^*} = 0.24$, 
$\varepsilon_{\Lambda \Xi^*} = 0.15$, 
$\varepsilon_{\Xi \Omega} = 0.08$, 
$\varepsilon_{\Sigma \Xi^*} = 0.08$,
$\varepsilon_{N \Sigma} = 0.05$, 
$\varepsilon_{\Lambda \Xi} = 0.05$, 
$\varepsilon_{N \Lambda} = 0.04$, 
$\varepsilon_{\D \Sigma^*} = 0.04$,
$\varepsilon_{\Sigma^* \Xi^*} = 0.04$, 
$\varepsilon_{\Xi^* \Omega} = 0.04$, 
and
$\varepsilon_{\Sigma \Xi} = 0.04$,
while for the 
$\D S = 1$ 
virtual transitions, 
the parameters are:
$\varepsilon_{\Lambda \Delta} = 0.04$,  
$\varepsilon_{\Xi \Sigma^*} = 0.04$, 
and
$\varepsilon_{\Sigma \Delta} = 0.04$.
For a majority of the strangeness changing transitions, 
the mass-splittings play little role in the 
$SU(2)$ 
expansion, 
i.e.~$\varepsilon_{BB'} \approx \varepsilon_{SU(2)}$. 
Despite the nearness of thresholds (compared to the kaon mass), 
the expansion parameters in 
$SU(2)$ 
are all better than the generic expansion parameter for 
$SU(3)$, 
$\varepsilon_{SU(3)} = m_\eta / M^{SU(3)} \sim 0.5$. 
Finally we remark that a perturbative treatment in 
$SU(2)$ 
excludes non-analytic pion mass dependence to describe the kaon threshold. 
For sufficiently small expansion parameters, 
the kaon threshold can be described by terms analytic in the pion mass-squared, 
but obviously non-analytic with respect to the strange quark mass.

\section{Effect of Kaon Thresholds}                                                                 %
\label{s:effect}                                                                                                   %

\subsection{Hyperon Masses}

The masses of spin-1/2 and spin-3/2 hyperons have been determined using 
$SU(2)$
chiral perturbation theory~\cite{Tiburzi:2008bk}. 
As the spin-3/2 resonances are closest to the kaon production threshold, 
we address how well the non-analyticities associated with the virtual 
process are described in the 
$SU(2)$ 
theory. 
Based upon our schematic arguments given above, 
we expect the virtual threshold to be well described by terms
non-analytic in the strange quark mass, 
but analytic in the pion mass-squared.

We can investigate the degree to which kaon thresholds 
affect hyperon masses by analyzing the associated kaon loop contributions. 
At leading-loop order, 
these arise from sunset diagrams. 
For the virtual process 
$B \to K B'$, 
the sunset diagram evaluates to
\begin{eqnarray}
\mathcal{F} (m_K^2, - \delta_{BB'}, \mu)
&=&
(m_K^2 - \delta_{BB'}^2)
\left[
\left( \d_{BB'}^2 - m_K^2 \right)^{1/2} 
\cL \big(m_K^2, - \d_{BB'} \big)
+
\d_{BB'} \log \frac{m_K^2}{\mu^2}
\right]
\notag \\
&& \phantom{space}
+
\frac{1}{2} \d_{BB'} \, m_K^2 \log \frac{m_K^2}{\mu^2}
\label{eq:F}
,\end{eqnarray}
up to overall group theory factors, axial couplings, and inverse powers of the chiral symmetry breaking scale. 
The dependence on 
$\mu$
is exactly cancelled by the scale dependence of 
local contributions to the hyperon mass 
which are at the same order in the chiral expansion. 
The logarithms appearing in Eq.~\eqref{eq:F} are straightforward to treat in the 
$SU(2)$
chiral expansion, as they are only functions of the kaon mass. 
One can use 
Eq.~\eqref{eq:kaon},
and expand in powers of the pion mass.
This part of the 
$SU(2)$ 
expansion is well behaved due to the size of the expansion parameter, $\varepsilon_{SU(2)}$.

To isolate the long-distance physics associated with the kaon threshold in the sunset diagram, 
we merely evaluate the function at the scale
$\mu = m_K$, 
which results in the function
\begin{equation} \label{eq:NoLog}
\mathcal{F} (m_K^2, - \delta_{BB'})
\equiv
\mathcal{F} (m_K^2, - \delta_{BB'}, \mu = m_K)
=
- 
(\d_{BB'}^2 - m_K^2)^{3/2} 
\cL \big(m_K^2, - \d_{BB'} \big)
.\end{equation} 
When one is near the threshold from above, 
$\d_{BB'} \gtrsim m_K$, 
this function has the behavior
\begin{equation}
\mathcal{F} (m_K^2, -\d_{BB'})
\to
2 \pi i (\d_{BB'}^2 - m_K^2)^{3/2}
+ \ldots \, \, 
,\end{equation}
which leads to the width for the decay process.
The functional form of the width is dictated by 
the available two-body phase space at threshold, 
and the requirement that the kaon and 
$B'$ 
baryon be in a relative 
$p$-wave.

%
\begin{figure}[t]
\begin{center}
\epsfig{file=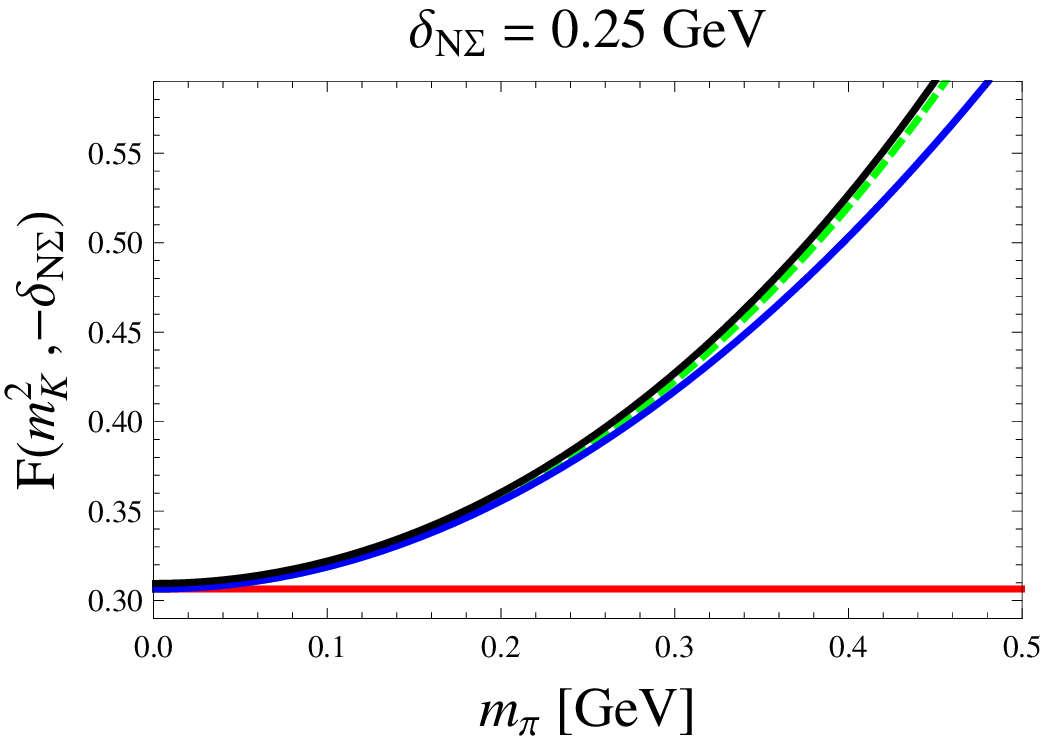,width=1.95in}
$\quad$
\epsfig{file=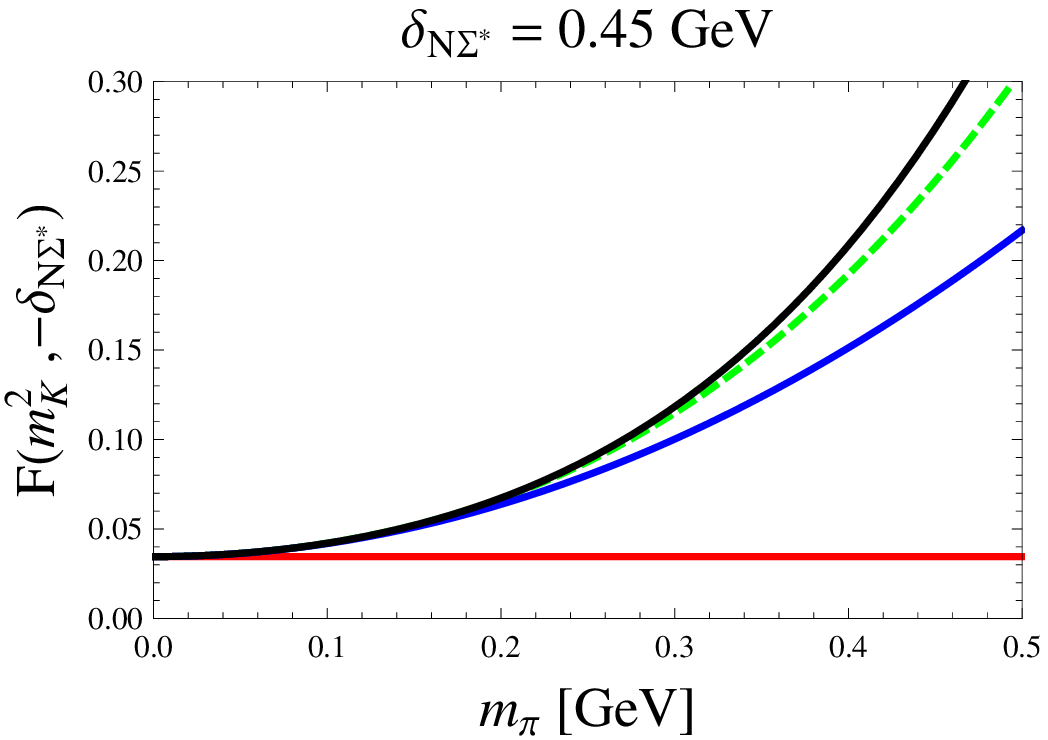,width=1.95in}
$\quad$
\epsfig{file=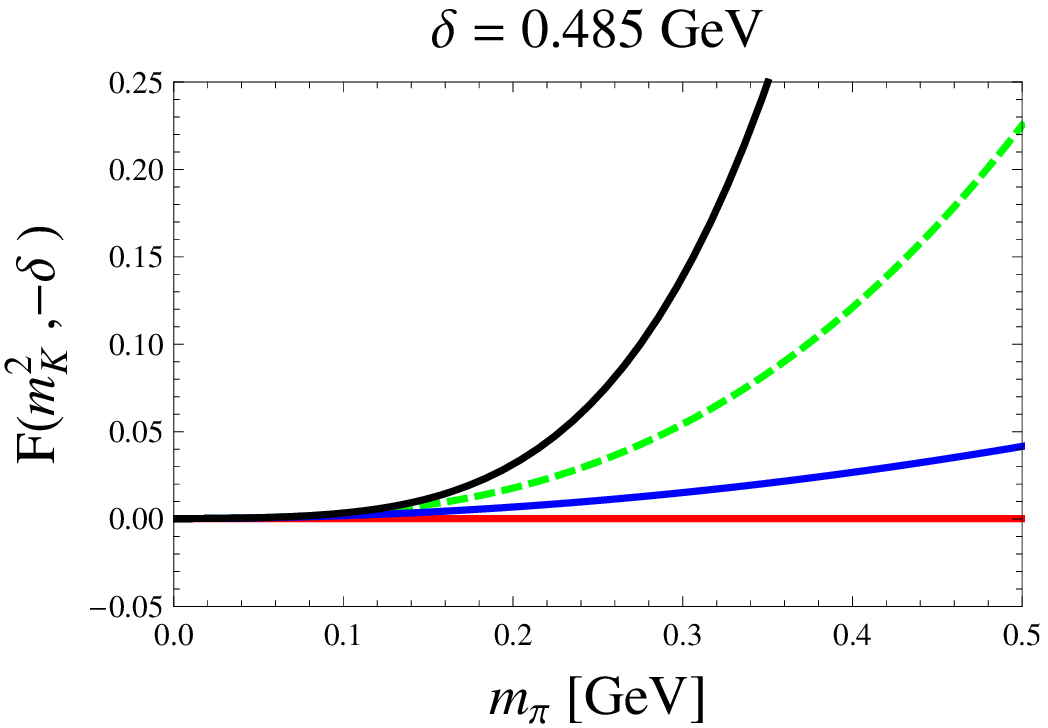,width=1.95in}
\caption{\label{f:expand} 
Virtual threshold contribution from the
$K$-$N$ 
sunset diagram for the $\Sigma$ (left), 
and 
$\Sigma^*$ (middle) baryon masses.
Plotted versus the pion mass and shown in dashed green 
is the non-analytic contribution 
$\mathcal{F}(m^2_K, - \delta_{BB'})$.
Also shown is the virtual contribution for a heavier external-state baryon (right)
with splitting $\d = 0.485 \, \texttt{GeV}$.  
Compared with these curves are three approximations that are analytic in the pion mass squared. 
The red curve is the zeroth-order approximation, 
$\cF^{(0)}$, 
while the blue curve also includes the first-order correction,
$m_\pi^2 \, \cF^{(2)}$, 
and finally the black curve includes all terms to 
$m_\pi^4$. 
Notice we show the same range, albeit shifted, in each plot. 
}
\end{center}
\end{figure}
%

For the sub-threshold case, 
the expansion of the function 
$\cF(m_K^2, - \d_{BB'})$
in 
$SU(2)$
chiral perturbation theory results in a perturbative series governed by 
$\varepsilon_{BB'}$
given in Eq.~\eqref{eq:discovery}. 
Specifically we have
\begin{equation} \label{eq:FExpand}
\cF(m_K^2, - \d_{BB'})
=
\cF^{(0)} 
+ 
C \,
m_\pi^2 \,
\cF^{(2)}
+
C^2 \, 
m_\pi^4 \,
\cF^{(4)}
+ 
\ldots
,\end{equation}
where the coefficients are implicitly functions of the strange quark mass and the baryon mass splitting. 
The first few coefficients are given by
\begin{eqnarray}
\cF^{(0)}
&=& 
\cF \left( [m_K^{SU(2)}]^2, - \d_{BB'} \right),
\notag \\
\cF^{(2)}
&=&
\frac{\d_{BB'} }{ 2 [m_K^{SU(2)}]^2 }
\left( \d_{BB'}^2 -[m_K^{SU(2)}]^2 \right)
+
\frac{3}{4}
\left( \frac{1}{ [m_K^{SU(2)}]^2 - \d_{BB'}^2} \right)
\cF^{(0)},
\notag \\
\cF^{(4)}
&=&
- \frac{\d_{BB'}}{8 [m_K^{SU(2)}]^4 }
\left( \d_{BB'}^2 + \frac{3}{2} [m_K^{SU(2)}]^2 \right)
+
\frac{3}{32} 
\left( \frac{1}{ [m_K^{SU(2)}]^2 - \d_{BB'}^2 }\right)^{2}
\cF^{(0)}
.\end{eqnarray}
Notice that by utilizing Eq.~\eqref{eq:discovery}, 
we have dropped terms of 
$\mathcal{O} ( \varepsilon_{BB'} \, \varepsilon_{SU(2)} )$.
For the case of near threshold processes, 
this approximation is legitimate because
$\varepsilon_{BB'} \, \varepsilon_{SU(2)} \ll  \varepsilon_{BB'}^2$. 
From these explicit terms in the expansion, 
we see that the virtual kaon threshold present in the sunset diagram
has been reduced to a sum of terms analytic in the pion mass squared, 
but non-analytic with respect to the strange quark mass.

To explore the 
$SU(2)$
expansion of kaon contributions to hyperon masses, 
we show the non-analytic contribution, 
Eq.~\eqref{eq:NoLog},
to the masses of
$\Sigma^*$ 
and 
$\Sigma$ 
baryons arising from virtual 
$K$-$N$ 
fluctuations in Figure~\ref{f:expand}. 
This result is compared with successive approximations
derived by expanding about the 
$SU(2)$ 
chiral limit, 
as in Eq.~\eqref{eq:FExpand}. 
We use the leading-order values of masses, 
and set 
$C=1$ 
to avoid estimating unknown low-energy constants. 
The plots show the non-analytic contribution 
associated with the virtual kaon threshold
can be captured in the two-flavor effective theory.
This is possible because the non-analyticities are 
dominated by the strange quark mass, 
whereas the lighter quark mass can be treated as a perturbation. 
Figure~\ref{f:expand} 
confirms that the expansion in terms of 
$\varepsilon_{BB'}$ 
in Eq.~(\ref{eq:expand}) 
is under control
for the range of values corresponding to the
$|\Delta S| = 1$ 
transitions, 
because in general  
$\varepsilon_{BB'} \leq \varepsilon_{N\Sigma^*}$.
The figure also depicts the case where the mass splitting has the value
$\d = 0.485 \, \texttt{GeV}$, 
which corresponds to an expansion parameter of size
$\varepsilon_{BB'} =  6.9$
at the physical pion mass.  
The series expansion in 
$m_\pi^2$
does not better approximate the non-analytic result with the addition of higher-order terms. 
As the series is in general asymptotic, 
the first term gives the best agreement when the expansion has broken down. 
For a fixed strange quark mass, 
there will always be a value of the pion mass above which the series breaks down. 
This value depends delicately on the size of the baryon mass splitting.

\subsection{Isovector Axial Charges}

Having explored the effect of the virtual kaon threshold on hyperon masses, 
we now turn to address the same effect on the hyperon isovector axial charges. 
The isovector axial charges of spin-1/2 hyperons have been determined using 
$SU(2)$
chiral perturbation theory~\cite{Jiang:2009sf}. 
This study was motived by the poor performance of 
$SU(3)$
chiral perturbation theory in describing lattice QCD data~\cite{Lin:2007ap}.  
The corresponding axial charges of spin-3/2 hyperons have not been studied in 
$SU(2)$ 
or 
$SU(3)$, 
with the exception of a large-$N_c$ analysis~\cite{FloresMendieta:2000mz}, 
and the axial charge of the delta resonances~\cite{Jiang:2008we}.
In the latter work, 
the delta axial charge was shown to exhibit strong non-analytic behavior with respect to the pion mass. 
The relatively large value of 
$g_{\D\D}$, 
or of its 
$SU(3)$ 
incarnation,
$\mathcal{H}$,
could undermine the chiral expansion of baryon properties. 
The commonly adopted value,
$g_{\D\D} \sim 2.2$, 
however, 
has only been inferred from chiral perturbation theory calculations.
Such calculations of 
$g_A$, 
or of 
$D$ 
and 
$F$ 
in 
$SU(3)$,
obtain the resonance axial coupling by neglecting local terms
which contribute at the same order in the expansion~\cite{Butler:1992pn,Bernard:1998gv}. 
With lattice QCD, 
it will be interesting to measure and study 
$SU(3)$
breaking in the axial charges of hyperon resonances.%
\footnote{
This will not be an easy task as the pion mass is lowered to the physical point.
Resonance properties can be studied from Euclidean space correlation functions through finite volume effects. 
Such studies are at an early stage%
~\cite{Gockeler:2008kc}, 
and have thus far focused on determining masses and widths of resonances. 
}
To this end, 
we analyze the behavior of kaon loops to determine whether an 
$SU(2)$ 
treatment for the resonances is feasible. 
Analyzing such contributions for the spin-1/2 hyperons, 
moreover,
justifies the findings in~\cite{Jiang:2009sf}, 
where it was argued that an 
$SU(2)$
treatment would better describe lattice data compared to 
$SU(3)$.

At leading loop order, 
one encounters a variety of diagrams in the evaluation of axial-vector
current matrix elements, 
for example, 
see~\cite{Jiang:2008aqa}.
The tadpole diagram with a kaon, 
of course, 
does not produce a threshold;
only the diagrams of sunset type contain the non-analyticities associated with kaon production.  
With a kaon loop, 
the general sunset diagram consists of 
a vertex for the process $B' \to K B$, 
followed by an axial current interaction  
$B \to B''$. 
For the isovector axial current, 
this is an isospin transition,
possibly also a transition from a spin-1/2 baryon to a spin-3/2 baryon or vice versa.
The remaining vertex encodes the process
$K B'' \to B'$. 
Evaluation of a loop diagram of this type produces terms proportional to the non-analytic function
\begin{equation} \label{eq:Ifunc}
\cI (m_K^2,  - \d_{BB'} , - \d_{B''B'}  , \mu )
= 
\frac{2}{3} \frac{1}{\d_{BB'} - \d_{B''B'}} 
\Big[ 
\cF(m_K^2, -\d_{BB'}, \mu)
-
\cF(m_K^2, -\d_{B''B'}, \mu)
\Big]
.\end{equation} 
Notice we have related this function to the non-analytic function arising in the mass sunset diagram.
This is possible because the product of the two intermediate-state baryon propagators can be written
as a difference of two terms having only single baryon propagators.

In 
$SU(2)$
chiral perturbation theory, 
the most subtle contributions to analyze arise from the external-state baryon 
$B'$
fluctuating into a kaon plus an intermediate-state baryon 
$B$ 
that is lighter than the 
$B'$. 
Let us focus on the 
$\S^*$ 
baryon as a concrete example for the worst-case scenario. 
Suppose that the first meson coupling produces a nucleon,
$\S^* \to K N$. 
The second meson coupling in the diagram depends on the action of the axial current insertion. 
There are two possible isovector axial-current insertions:
baryon spin changing, and baryon spin conserving. 
For the baryon spin-changing current, 
the nucleon transitions to a delta, 
$N \to \D$
with an axial coupling proportional to $g_{\D N}$.
The second meson coupling is required to be 
$K \D \to \S^*$, 
and the corresponding diagram is proportional to the function
$\mathcal{I} ( m_K^2, - \d_{N\S^*}, - \d_{\D \S^*}, \mu)$. 
By virtue of the algebraic simplification made in Eq.~\eqref{eq:Ifunc}, 
this contribution can be expressed in terms of 
$\cF(m_K^2, - \d_{N\S^*}, \mu)$ 
and 
$\cF(m_K^2, - \d_{\D \S^*}, \mu)$. 
The 
$SU(2)$
expansion of this function has been detailed above in the context of hyperon masses. 
We thus conclude that the non-analyticites present in the sunset diagram with the axial transition
$\S^* \to K N \to K \D \to \S^*$
can be described in an
$SU(2)$
chiral expansion.

%
\begin{figure}[t]
\begin{center}
\epsfig{file=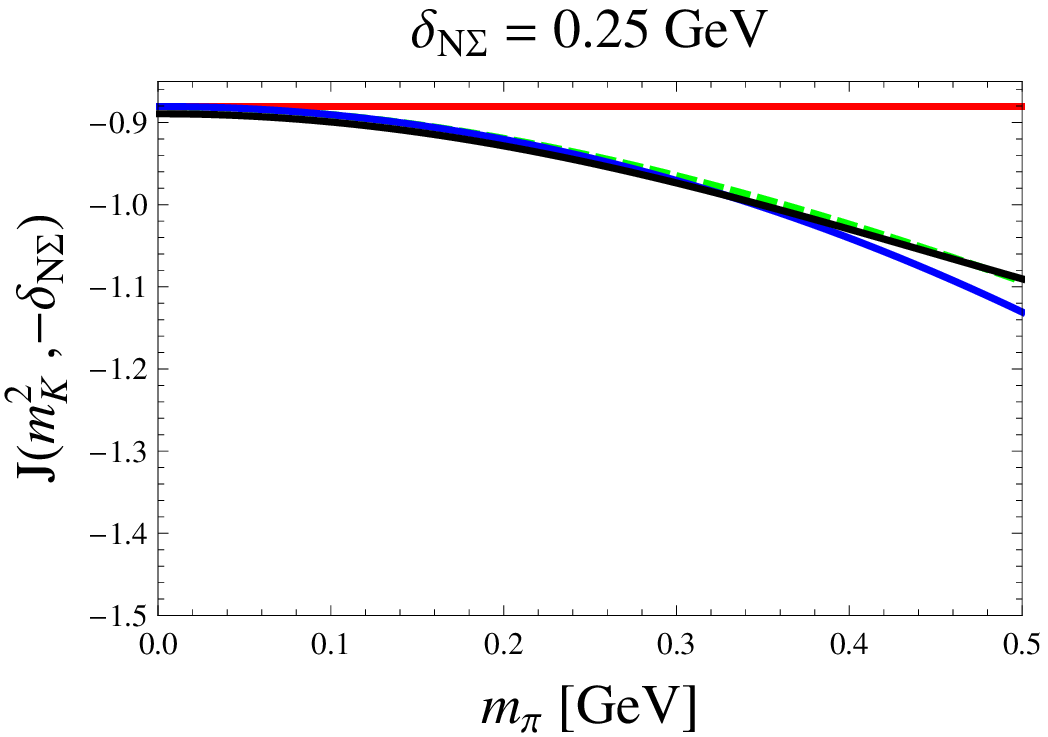,width=1.95in}
$\quad$
\epsfig{file=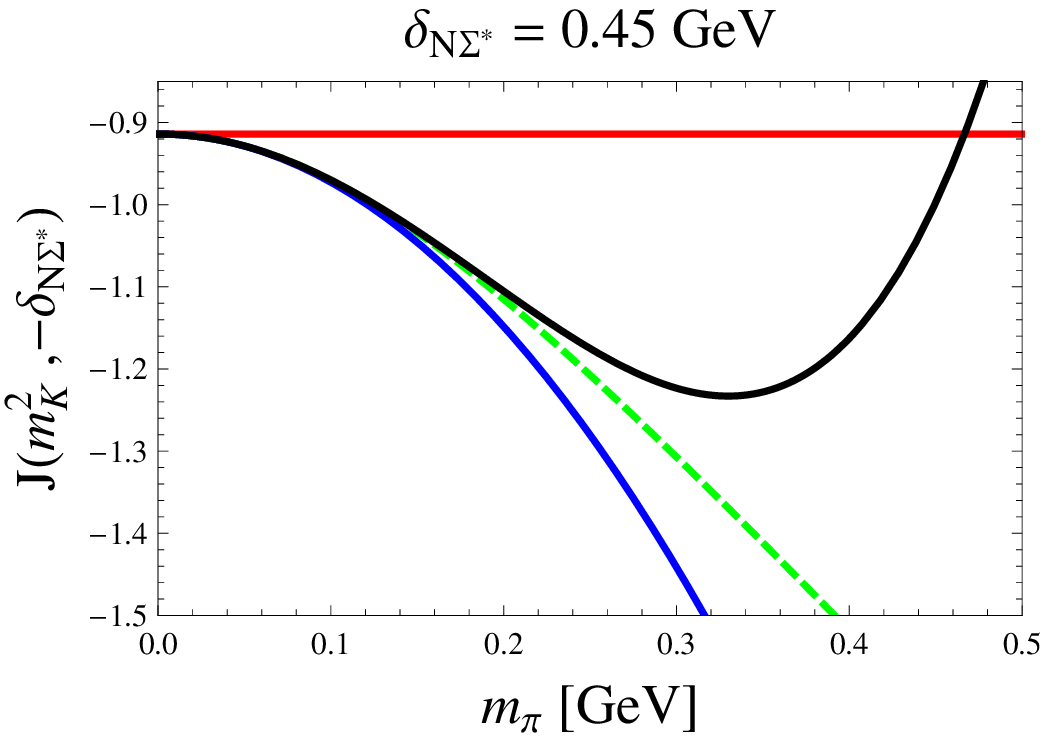,width=1.95in}
$\quad$
\epsfig{file=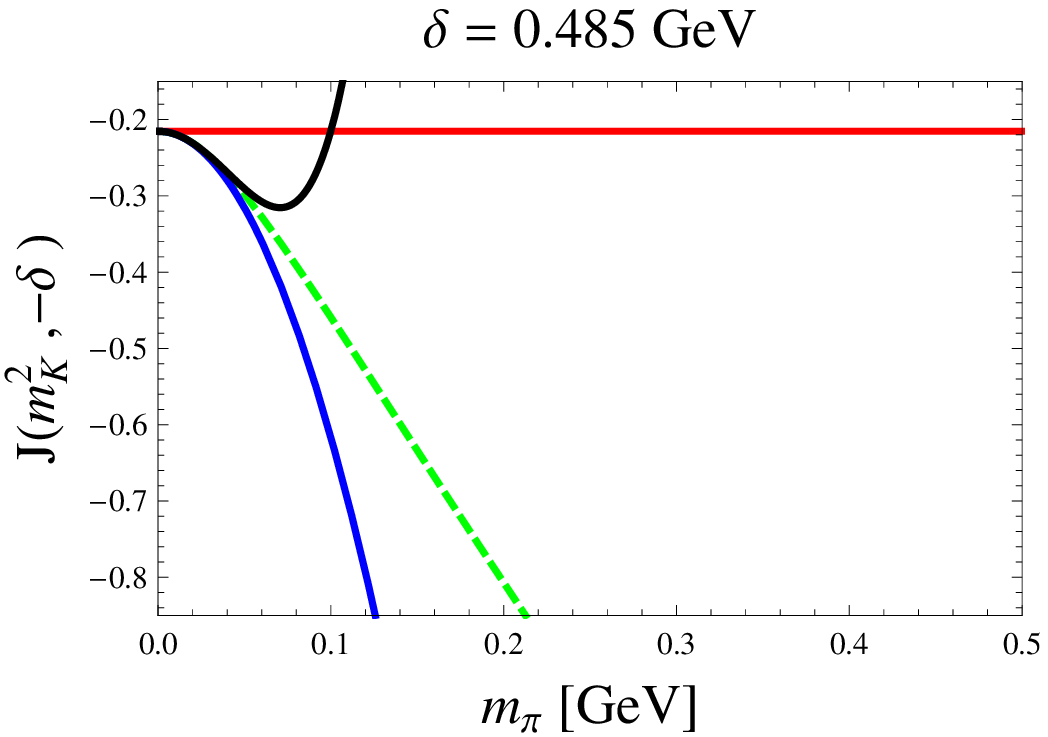,width=1.95in}
\caption{\label{f:axial} 
Virtual threshold contribution from the kaon sunset diagram with intermediate-state isovector axial transition
$N \to N$ 
for the $\Sigma$ (left), 
and 
$\Sigma^*$ (middle) baryons.
Plotted versus the pion mass and shown in dashed green 
is the non-analytic contribution 
$\mathcal{J}(m^2_K, - \delta_{BB'})$.
Also shown is the virtual contribution for a heavier external-state baryon (right)
with splitting $\d = 0.485 \, \texttt{GeV}$.  
Compared with these curves are three approximations that are analytic in the pion mass squared. 
The red curve is the zeroth-order approximation, 
$\cJ^{(0)}$, 
while the blue curve also includes the first-order correction,
$m_\pi^2 \, \cJ^{(2)}$, 
and finally the black curve includes all terms to 
$m_\pi^4$. 
Notice we show the same range, albeit shifted, in each plot. 
}
\end{center}
\end{figure}
%

The spin-conserving axial current requires a more detailed analysis. 
For our example of the 
$\S^*$ 
baryon, 
the intermediate state 
$N$ 
then makes an isovector transition with coupling 
$g_A$, 
and the final vertex describes the process
$K N \to \S^*$. 
Such kaon sunset diagrams are proportional to the non-analytic function
\begin{equation}
\cJ (m_K^2, - \d_{BB'}, \mu)
= 
\mathcal{I} (m_K^2, - \d_{BB'}, -\d_{BB'}, \mu)
,\end{equation}
with 
$\d_{BB'} = \d_{N\S^*}$ 
for the specific example of the 
$\S^*$. 
Taken at the scale
$\mu = m_K$, 
the function 
$\cJ$
contains only long-distance contributions associated with kaon production;
explicitly we have
\begin{equation} \label{eq:Jfunc}
\cJ (m_K^2, - \d_{BB'} )
\equiv
\cJ ( m_K^2, -\d_{BB'}, \mu = m_K )
= 
- 2 \d_{BB'} 
( \d_{BB'}^2 - m_K^2)^{1/2} 
\cL \big(m_K^2, - \d_{BB'} \big) 
.\end{equation}
At threshold, 
this function is proportional to the available phase space. 
Appealing to a perturbative
$SU(2)$
expansion, 
we write
\begin{equation}
\cJ (m_K^2, - \d_{BB'} )
=
\cJ^{(0)}
+ 
C \,
m_\pi^2 \, \cJ^{(2)}
+
C^2 \,
m_\pi^4 \, \cJ^{(4)}
+ 
\ldots
,\end{equation}
where the coefficients are implicitly non-analytic functions of the strange quark mass
and baryon mass splitting. 
Omitted terms are proportional to higher powers of the pion mass-squared. 
The first three coefficients in the expansion are given by
\begin{eqnarray}
\cJ^{(0)}
&=& 
\cJ 
\left(  [m_K^{SU(2)}]^2, - \d_{BB'}
\right)
,\notag \\
\cJ^{(2)} 
&=&
\frac{\d_{BB'}^2}{[m_K^{SU(2)}]^2}
+
\frac{1}{4} 
\left(
\frac{1}{[m_K^{SU(2)}]^2 - \d_{BB'}^2 }
\right)
\cJ^{(0)}
,\notag \\
\cJ^{(4)}
&=&
-
\frac{\d_{BB'}^2}{8 [m_K^{SU(2)}]^4}
\left(
1
-
\frac{\d_{BB'}^2}{[m_K^{SU(2)}]^2 -\d_{BB'}^2 }
\right)
-
\frac{1}{32} 
\left(
\frac{1}{ [m_K^{SU(2)}]^2 - \d_{BB'}^2} \right)^2
\cJ^{(0)}.
\end{eqnarray}
Because our interests lie in near threshold virtual processes, 
we have again utilized Eq.~\eqref{eq:discovery}, 
and dropped terms of
$\mathcal{O} ( \varepsilon_{BB'} \, \varepsilon_{SU(2)} )$.

To explore the
$SU(2)$
expansion of kaon contributions to axial charges arising from sunset diagrams with processes of the form 
$B' \to K B \to K B \to B'$, 
we plot the non-analytic contribution, 
Eq.~\eqref{eq:Jfunc}, 
as a function of the pion mass in Figure~\ref{f:axial}.
We specialize to the case of the 
$\S$ 
and 
$\S^*$ baryons, 
for which the splittings 
$\d_{N\S}$ 
and 
$\d_{N\S^*}$
are relevant, 
respectively.
To avoid uncertainties with parameter values, 
we utilize leading-order estimates, 
and thereby take 
$C=1$. 
We also consider the case of a fictitious external-state baryon which has a mass splitting with the nucleon of
$\d = 0.485 \, \texttt{GeV}$.  
In the case of the 
$\S$ 
and
$\S^*$,
the plots show the non-analytic virtual kaon contribution can be captured 
by terms in the effective theory that are analytic in the pion mass-squared.
These particular contributions, 
however, 
exhibit more sensitivity to the virtual threshold compared to contributions to the mass.
This sensitivity can be easily accounted for by studying the behavior at threshold:
the 
$\cF$-function 
vanishes with the third power of the available energy, 
while the 
$\cJ$-function 
only vanishes with the first power. 
Consequently the range of pion masses for which expansion is viable is more limited. 
For the 
$\S^*$, 
the expansion becomes unreliable past 
$m_\pi \sim 0.3 \, \texttt{GeV}$. 
For the fictitious heavier baryon, 
the expansion has broken down even at the physical pion mass, 
where the expansion parameter has the value
$\varepsilon_{BB'} = 6.9$. 
There is only a narrow range of pion masses about the 
$SU(2)$ 
chiral limit for which the expansion at 
$\d = 0.485 \, \texttt{GeV}$
exhibits convergence.

\section{Summary}                                                                                           %
\label{summy}                                                                                                   %

Above we explore the effect of kaon contributions on the properties of hyperons. 
Strangeness changing fluctuations allow a hyperon to make virtual transitions to kaons and baryons of smaller masses. 
Because some of these processes are not considerably far from the kaon production threshold, 
$\d_{BB'} \sim m_K$,
one requires non-analytic behavior with respect to the kaon mass-squared to describe such fluctuations. 
This can be accomplished with 
$SU(3)$
chiral perturbation theory at the price of a rather large expansion parameter,
$\varepsilon_{SU(3)} =  m_\eta / M^{SU(3)} \sim 0.5$.

To improve the convergence of the chiral expansion, 
one can alternately formulate theories of hyperons using an expansion about the
$SU(2)$ 
chiral limit.
The presence of kaon sub-thresholds na\"ively seems to complicate an 
$SU(2)$
description of hyperon properties, 
because explicit kaon contributions are absent. 
We show, 
however, 
that certain hyperon observables are amenable to an 
$SU(2)$ 
treatment. 
In the 
$SU(2)$
expansion, 
the relevant expansion parameter describing the kaon threshold is not
$\d_{BB'} / m_K$, 
but rather
$\varepsilon_{BB'}$
given in Eq.~\eqref{eq:discovery}.
For the most troublesome cases, 
the expansion parameters take on values smaller than  
$\varepsilon_{SU(3)}$. 
This remains true when higher-order corrections to the 
$SU(2)$ 
expansion parameters are estimated, 
although one requires higher-precision lattice data than currently available to arrive at a definitive  conclusion.

For hyperon masses and isovector axial charges, 
we find that non-analyticities associated with the kaon threshold in sunset diagrams
can be described in two-flavor chiral perturbation theory. 
While the two-flavor expansion of these thresholds contains only terms 
that are analytic in the pion mass-squared, 
the coefficients of such terms are non-analytic functions of the strange quark mass and baryon mass splittings. 
Certain contributions to hyperon axial charges exhibit greater sensitivity to the kaon threshold than others. 
This sensitivity arises from the behavior of the non-analytic contributions as the threshold is approached:
the slower the function vanishes at threshold, the more sensitive to the kaon threshold. 
While hyperon masses and isovector axial charges appear amenable to 
$SU(2)$ 
chiral perturbation theory, 
our observations do not generalize to all observables. 
In fact, 
our analysis shows a limitation of the two-flavor theory:
observables that become singular at the kaon threshold will not be well described by an expansion in the pion mass-squared.

Finally we remark that potential problems with kaon sub-thresholds
are only relevant for a description of hyperons explicitly including the spin-3/2
degrees of freedom.
One can thus attempt to dodge the issue by restricting the theory to only 
spin-1/2 states, and integrating out the virtual spin-3/2 fluctuations. 
The resulting theory is governed by an expansion parameter
$\varepsilon_{B^*} \sim m_\pi / \D_{BB^*}$, 
where 
$\D_{BB*}$ is the mass splitting between the spin-3/2, 
$B^*$, 
and spin-1/2, 
$B$, 
hyperon multiplets. 
This approach is less advantageous compared to the nucleon sector, 
for example, 
in the cascade sector at the physical pion mass one has 
$\varepsilon_{\Xi^*} \approx 2/3$. 
In the extrapolation of lattice data, 
moreover, 
one often has 
$\varepsilon_{B^*} \gtrsim 1$
which often necessitates the inclusion of spin-3/2 multiplets. 
The study of inelastic contributions to other observables is certainly needed to ascertain in which cases a two-flavor expansion is valid.  
Further, the utility of an 
$SU(2)$ 
treatment of hyperons, 
with the significant growth of LECs, 
probably requires the aid of lattice QCD calculations to determine all these unknown parameters.
Ultimately lattice QCD data will enable us to determine when the 
$SU(2)$ 
theory of hyperons is an effective one.


\begin{acknowledgments}
This work was supported by the
U.S.~Department of Energy, 
under
Grant Nos.%
~DE-FG02-94ER-40818 (F.-J.J.),
~DE-FG02-93ER-40762 (B.C.T.), and
~DE-FG02-07ER-41527 (A.W.-L.).
Additional support provided by the
Taiwan National Center for Theoretical Sciences, 
North; 
and the
Taiwan National Science Council
(F.-J.J.).
\end{acknowledgments}


\appendix

\section*{Higher-Order Corrections} %

Our assessment of 
$SU(2)$
chiral perturbation theory for hyperons relies on estimating
the kaon mass and baryon mass splittings in the 
$SU(2)$ 
chiral limit. 
The expansion parameters underlying 
$SU(2)$ 
depend quite sensitively on these masses. 
For example, 
reducing  
$m_K^{SU(2)}$ 
by
$10 \%$
from that in Eq.~\eqref{eq:kaon2} 
shows that an expansion in 
$\varepsilon_{N \Sigma^*}$ 
is ill-fated. 
At this value of 
$m_K^{SU(2)}$, 
the expansion parameter is negative indicating that we have passed through the pole in 
Eq.~\eqref{eq:discovery} 
by lowering  
$m_K^{SU(2)}$. 
To further assess the convergence of
$SU(2)$,
we address the impact of next-to-leading order corrections.

Using 
$SU(3)$
chiral perturbation theory at next-to-leading order%
~\cite{Gasser:1984gg}, 
the 
$SU(2)$ chiral limit mass of the kaon can be written in the form:
$[m_K^{SU(2)}]^2
= 
m_K^2 
-
\frac{1}{2} C \, m_\pi^2$, 
with 
\begin{equation}
C
= 
1
+ 
32 \frac{m^2_{K}}{f^2}
\left(
\frac{  \log \frac{4 m^2_K }{ 3 \mu^2}  + \frac{1}{4} }{288 \pi^2}
+ 
2 L_8 (\mu) - L_5(\mu) 
+ 
2  
\left[ 
2 L_6(\mu) - L_4(\mu) 
\right] 
\right)
,\end{equation}
where we have dropped terms that behave as
$m_\pi^4$
because these are suppressed by a relative factor of
$\varepsilon_{SU(2)}$. 
To determine 
$m_K^{SU(2)}$
using this next-to-leading order expression,
we must rely on values for the low-energy constants. 
In Table~\ref{t:values}, 
we list estimates of these parameters and their sources.
Although there is considerable spread in values for the low-energy constants, 
the various sources produce the same kaon mass in the chiral limit
to about 
$1 \%$. 
The size of the next-to-leading order correction to  
$C$ 
is inline with expectations, 
but the value varies over the data sets by 
$\pm 20 \%$. 
Thus we have
$C = 1.0(2)$,
and adopt the central value for all estimates.  
Due to the pole present in 
$\varepsilon_{BB'}$, 
it is comparatively more important to improve the estimate of the denominator than the numerator.

%
  \begin{table}
  \begin{center}
     \caption{Values of low-energy constants taken from lattice QCD and phenomenology, 
     along with the resulting estimates of 
     $m_K^{SU(2)}$, 
     and the expansion parameters  
     $\varepsilon_{N\Sigma^*}$, 
     and 
     $\varepsilon_{\Lambda \Xi^*}$ using Eq.~\eqref{eq:discovery}.
     Gasser-Leutwyler parameters, 
     denoted by 
     $L_i$, 
     are quoted in units of 
     $10^{-4}$
     at the renormalization scale 
     $\mu = 0.770 \, \texttt{GeV}$. 
     As we are unable to take into account correlations among parameters, 
     we do not cite uncertainties. 
     }
    \smallskip
   \begin{tabular}{c|cc|ccc}
    Source & 
    $ \quad 2 L_6 - L_4 $ \quad & 
    $ \quad 2 L_8 - L_5  $ \quad &
    $ \quad m_K^{SU(2)} [ \texttt{GeV} ] \quad$ &  
    $\quad \varepsilon_{N\Sigma^*} \quad $ &
    $\quad \varepsilon_{\Lambda \Xi^*} \quad$ \\
    \hline
    \hline
   RBC/UKQCD~\cite{Allton:2008pn}  &
    $0.0$ & 
    $\phantom{-}2.4$ & 
    $0.486$ & 
    $0.33$ & 
    $0.17$ \\
    \hline
    2007 MILC~\cite{Bernard:2007ps}   &
    $3.4$ & 
    $\phantom{-}2.6$ & 
    $0.483$ &
    $0.37$ &
    $0.18 $\\
    \hline   
    2009 MILC Lattice~\cite{Bazavov:2009tw}   &
    $1.0$ & 
    $-1.2$ & 
    $0.487$ &
    $0.33$ &
    $0.16$    \\
    \hline   
    Phenomenology ``Main Fit''~\cite{Amoros:2001cp} &
    $\equiv 0$ &
    $\phantom{-}3.3$ & 
    $0.486$  &
    $0.34$ &
    $0.17$ \\
        \hline
    Phenomenology ``Fit D''~\cite{Bijnens:2007si} &
    $\equiv -2.0$ &
    $\phantom{-}2.0$ & 
    $0.488$ &
    $0.31$ &
    $0.16$ \\
    \hline
       \hline
       \end{tabular}
  \label{t:values}
  \end{center}
 \end{table}
%

For the baryon masses, 
we utilize the expansion about the 
$SU(2)$ 
chiral limit
\begin{equation}
M_B = M_B^{SU(2)} + \frac{\s_B}{4 \pi f} m_\pi^2 + \ldots \, \,
.\end{equation}
Here $f = 0.132 \, \texttt{GeV}$ is the pion decay constant, 
which is our conventional choice to make the low-energy constant dimensionless. 
Knowledge of the physical baryon masses, 
$M_B$,
and the 
$\sigma_B$
parameters enables us to determine the 
$SU(2)$
chiral limit value of the mass splittings, namely
\begin{equation}
\delta_{BB'}
=
M_{B'}^{SU(2)} -  M_{B}^{SU(2)}
=
M_{B'} -  M_{B} 
+ 
\frac{\sigma_{B'} - \sigma_B}{ 4 \pi f} m_\pi^2
+ \ldots \,
.\end{equation}
For estimates of the low-energy parameters, 
$\sigma_B$,  
we use those in~\cite{Tiburzi:2008bk} for the spin-1/2 baryons, 
and the procedure of~\cite{Tiburzi:2008bk} to estimate those for the spin-3/2 baryons 
using lattice data from~\cite{WalkerLoud:2008bp}.
For the two largest 
$|\D S| = 1$ 
baryon mass splittings, 
we need the values
$\sigma_N = 1.8(4)$,
$\sigma_\Lambda = 1.2(2)$,
$\sigma_{\S^*} = 0.75(15)$,
and
$\sigma_{\Xi^*} = 0.52(10)$.
The uncertainties have been somewhat arbitrarily assigned at 
$20 \%$, 
and are due to the
$SU(2)$
chiral extrapolation. 
From these values of 
$\sigma$-parameters
and the physical baryon masses, 
we arrive at the two largest
$SU(2)$
chiral limit baryon mass splittings:
\begin{eqnarray}
\delta_{N \Sigma^*}
&=& 0.457(4) \, \texttt{GeV}, \qquad 
\delta_{\Lambda \Xi^*}
= 0.426 (3)\, \texttt{GeV}
\label{eq:2splits}
.\end{eqnarray}
These values are only slightly larger than the physical splittings, 
because there is partial cancelation in differences of the
$m_\pi^2$ 
corrections.

Combining the chiral limit value of the kaon mass and baryon mass splittings, 
we can estimate the 
$SU(2)$
expansion parameters that govern the description of kaon thresholds beyond leading order. 
Values for 
$\varepsilon_{N\Sigma^*}$, 
and 
$\varepsilon_{\Lambda \Xi^*}$ 
derived using Eq.~\eqref{eq:discovery}
also appear in Table~\ref{t:values}. 
The values derived using lattice QCD input and phenomenology suggest that the leading-order expansion parameters 
given in Sec.~\ref{s:KPT}
have been underestimated. 
More precise determination requires lattice QCD values of
$SU(2)$
chiral limit masses at the level of a few 
$\texttt{MeV}$. 
Having considered the two worst possible baryon transitions, 
however,
we still expect the
$SU(2)$ 
chiral expansion to provide a good description of kaon threshold contributions to hyperon masses and isovector axial charges considered above.

\bibliography{hb}

\end{document}